\documentclass[preprint,
showkeys,
aps,
]{revtex4-1}

\usepackage[english]{babel}
\usepackage{graphicx}
\usepackage{dcolumn}
\usepackage{bm}

\usepackage[colorlinks=true,urlcolor=blue,citecolor=blue,linkcolor=red]{hyperref}
\usepackage{booktabs}
\usepackage{amsmath}
\usepackage{amssymb}
\usepackage{braket}
\usepackage[dvipsnames]{xcolor}
\usepackage{ulem}

\begin{document}

\title[Interacting quantum mixtures]{Interacting quantum mixtures for precision atom interferometry}

\author{Robin Corgier$^{1,2,}$\footnote[1]{Current address: Quantum Science and Technology Arcetri INO-CNR, Largo Enrico Fermi 2, 50125 Firenze, Italy}$^,$\footnote[2]{robin.corgier@gmail.com}, Sina Loriani$^1$, Holger Ahlers$^1$, Katerine Posso-Trujillo$^1$, Christian Schubert$^1$, Ernst M. Rasel$^1$, Eric Charron$^2$ and Naceur Gaaloul$^{1,}$\footnote[3]{gaaloul@iqo.uni-hannover.de}}

\address{$^1$ Institut f\"ur Quantenoptik, Leibniz Universit\"at Hannover, Welfengarten 1, 30167 Hannover, Germany}
\address{$^2$ Universit\'e Paris-Saclay, CNRS, Institut des Sciences Mol\'eculaires d'Orsay, 91405 Orsay, France}

\begin{abstract}
We present a source engineering concept for a binary quantum mixture suitable as input for differential, precision atom interferometry with drift times of several seconds. To solve the non-linear dynamics of the mixture, we develop a set of scaling approach equations and verify their validity contrasting it to the one of a system of coupled Gross-Pitaevskii equations. 

This scaling approach is a generalization of the standard approach commonly used for single species. Its validity range is discussed with respect to intra- and inter-species interaction regimes. We propose a multi-stage, non-linear atomic lens sequence to simultaneously create dual ensembles with ultra-slow kinetic expansion energies, below 15\,pK. Our scheme has the advantage of mitigating wave front aberrations, a leading systematic effect in precision atom interferometry.
\end{abstract}

\keywords{Bose–Einstein condensate, quantum mixtures, atom interferometry, scaling approach, equivalence principle, interacting quantum gases.}

\maketitle

\section{Introduction}
\label{sec:Intro}

The high precision of atom interferometry-based sensors makes them an exquisite tool for performing tests of fundamental theories~\cite{dimopoulos_testing_2007, dimopoulos_general_2008, dimopoulos_atomic_2008, fixler_atom_2007, lamporesi_determination_2008, bouchendira_new_2011} as well as for metrology~\cite{PRL_Pezze_2009,Nature_Riedel_2010,Science_Lucke_2011,RevModPhy_Pezze_2018}, geodesy or inertial navigation~\cite{Geiger2020arxiv}. One timely challenge is to test the weak equivalence principle (WEP) or universality of free fall (UFF)~\cite{damour_testing_1996} by tracking the acceleration of two different test masses in free fall using matter-wave interferometry~\cite{berman_atom_1997, cronin_optics_2009}. This experimental test~\cite{will_confrontation_2010} is important in the context of Grand Unification theory~\cite{kiefer_quantum_2007} to falsify some of the competing models, which predict a violation of the UFF at different levels~\cite{damour_violations_2002,sandvik_simple_2002,wetterich_probing_2003}.  EP tests are parametrised by the Eötvös ratio $\eta$, which is the relative acceleration of the test masses divided by their average acceleration in the same gravitational field.
The simultaneous operation of a dual-species (or isotopes) atom interferometer (AI) was proposed and expected~\cite{fray_atomic_2004, dimopoulos_testing_2007} to perform a UFF test with a target performance that would exceed the best reported measurements using classical test masses as torsion balances at $\eta=1.8\times10^{-13}$~\cite{schlamminger_test_2008}, Laser Lunar Ranging at $\eta=1.4\times10^{-13}$~\cite{williams_progress_2004} or the space mission MICROSCOPE at $\eta=1.3\times10^{-14}$~\cite{PRLTouboul2017}.

The sensitivity of an atomic inertial sensor scales quadratically with the time spent by the atoms inside the interferometer~\cite{berman_atom_1997}, limiting ground-bound UFF tests~\cite{bonnin_simultaneous_2013, tino_testing_2013,PRLSchlippert2014,Asenbaum2020arxiv,Albers2020arxiv} and motivating the drive for long free expansions. Several platforms are therefore considered for such an increase, such as droptowers~\cite{van_zoest_bose-einstein_2010, rudolph_degenerate_2011}, fountains~\cite{muller_atom-interferometry_2008, Zhou2011, VLBAI,Asenbaum2020arxiv}, parabolic flights~\cite{geiger_detecting_2011}, the international space station~\cite{CALwebsite,Fry_2019} and spacecraft~\cite{tino_precision_2013, NatureBecker2018, ASPALTSCHUL2015}. To match the UFF performance of classical tests, the interferometry time has to reach the second regime~\cite{Gaaloul2014}. Combined with a satellite operation, a performance in the range of $\eta=10^{-15}-10^{-17}$~\cite{Battelier2019} can be made possible. To be able to observe the atomic ensemble after several seconds, one uses ultra-cold degenerate gases that exhibit ultra-slow expansion and unique coherence properties~\cite{lamine_ultimate_2006}. Recently, AIs fed with single-species Bose-Einstein condensates (BEC)~\cite{cornell_nobel_2002, ketterle_nobel_2002} were operated in these long-time regimes~\cite{PRLMuntinga2013, dickerson_multiaxis_2013} by taking advantage of the delta-kick collimation (DKC) technique~\cite{chu_proposal_1986, ammann_delta_1997, morinaga_manipulation_1999,PRLMuntinga2013, PRLKovachy2015} since a simple free expansion would quickly lead to low atomic densities that could be below the density threshold for detection~\cite{berman_atom_1997}. These demonstrations point towards a high-accuracy UFF test when combined with a second species in a differential atom interferometry measurement. The source of such an interferometer would naturally be a quantum degenerate atomic mixture. Such an input state would allow to go beyond the current recent performance of $\eta=10^{-12}$~\cite{Asenbaum2020arxiv} where a binary non-condensed source was used taking advantage of the reduced systematic effects associated to quantum gases.

In this paper, we theoretically study the use of these dual sources from the specific angle of their appropriateness in an atom interferometric accurate measurement. Quantum mixtures of ultra-cold gases received a surge of theoretical~\cite{PRLHo1996, pu_properties_1998, PRLEsry1997, JPB_Trippenbach_2000} and experimental~\cite{hall_dynamics_1998, myatt_production_1997, PRLFerrari2002, PPRLMcNamara2006, mertes_nonequilibrium_2007, PRLPapp2008,Thalhammer_2008, PRLJorgensen2016, PRA_Burchianti_2018, PRASchulze2018} interest since the early years of BEC manipulation revealing extremely rich and interesting physics. Nevertheless, studies of degenerate mixtures as an appropriate source satisfying the requirements of a differential AI in the conditions described above have not been considered to our knowledge. Neither the size, the density shape control, the common collimation of the two species, nor the AI-relevant systematics are specifically reported. In addition to covering these aspects, we have paid particular attention to the effects of interactions, which cannot be neglected for the desired number of atoms in an AI with high sensitivity, \textit{i.e.} about $10^5$ to $10^6$ atoms in each condensate.

We illustrate our theoretical approach with the study of a degenerate mixture of $^{87}$Rb and $^{41}$K recently proposed for a cutting-edge UFF test~\cite{CQGAguilera2014,Battelier2019}. 
This choice is justified first by the possibility to tune the interaction between the two species, where the presence of two Feshbach resonances below 100\,G has been demonstrated~\cite{PRLFerrari2002}. 
A second motivation is related to the miscibility of the degenerate mixture with both species sharing the same center of mass.
This feature is of particular interest since an offset between the center-of-mass of the two species coupled to gravity and/or magnetic field gradients can lead to large detrimental systematic effects~\cite{Hogan08,Schubert2013, CQGAguilera2014, ASPALTSCHUL2015}.
Third, it is essential to explore various combinations of species in order to put bounds on different UFF violation models~\cite{Kostelecky_2011,PRDHohensee2010,CQGDamour2012}, which parametrize composition-dependent couplings to fifth forces, dark matter, etc. 
Depending on the model under consideration, it is advantageous to choose different species over different isotopes of the same species, since violation parameters scale favourably with the baryon numbers. Here, by means of a coupled scaling approach derived for miscible mixtures and verified by Gross-Pitaevskii equations, we found a suitable regime for an AI operating during more than 10\,s using a two-component Rb-K BEC source. 
This is expected to allow for an accuracy of UFF tests of few parts in $10^{-15}$ for the E\"otv\"os ratio~\cite{Schubert2013, CQGAguilera2014, ASPALTSCHUL2015}. 
Mitigation strategies for leading systematic errors as the wave front aberrations that would limit the accuracy of UFF tests are also proposed. We show that the regime of extremely long times (several seconds) and ultra-slow kinetic expansion speeds (around 10\,$\mathrm{\mu}$m.s$^{-1}$) is accessible within current experimental capabilities.

This paper is structured as follows: In Sec.\,\ref{sec:scheme} we briefly recall the salient features of the proposed scheme. Sec.\,\ref{sec:theory} presents the theoretical tools used to track the dynamics of the quantum mixture. In Sec.\,\ref{sec:results}, we illustrate the preparation scheme of the mixture, its rich ground state properties and the long-time dynamics observed. In Sec.\,\ref{sec:high-precision} we discuss aspects that are important when using the prepared mixture as an input of a differential AI, in particular the matching of expansion rates. Conclusions and perspectives are presented in Sec.\,\ref{sec:conclusion}.

\section{System considered and proposed sequence}
\label{sec:scheme}

In spite of the high number of characteristic parameters (three scattering lengths, two different atom numbers, different trap characteristics, etc..), two-component mixtures could be classified in two general categories depending whether their spatial density distributions form  symmetric or asymmetric patterns, often referred to as miscible and immiscible states, respectively.~\cite{PRLHo1996, PRLEsry1997, JPB_Trippenbach_2000}. The system that we shall consider as a study case in this paper consists of a BEC binary mixture of $^{87}$Rb and $^{41}$K. A two-component BEC of this kind was first produced in the pioneering experiments of Ref.~\cite{PhysRevA.98.063616}. To be close to state-of-the-art realisations, we consider 10$^5$ atoms in each BEC. Since at vanishing ambient magnetic field, the (positive) inter-species s-wave scattering length is larger than the two (positive) intra-species ones, the two BECs repel each other for such high atom numbers. This leads to a spatial separation of the two BECs, even though they share the same center-of-mass. One of the two species is located in the center of the trap, surrounded by the second one in an onion-like shape. In presence of a magnetic field, the interaction between these two species can be tuned thanks to Feshbach resonances~\cite{Thalhammer_2008,Chin_2010}. For a particular value of this magnetic field, the inter-species interaction vanishes and the two BECs feature a large overlap region. If the trap is released, the atomic clouds expand freely in the Feshbach magnetic field. A delta-kick collimation (DKC) stage can follow, during which the initial trap is briefly switched-on again. This pulsed trap strategy is intended to remove a substantial part of the kinetic energy from the expanding gas~\cite{chu_proposal_1986, ammann_delta_1997, morinaga_manipulation_1999, PRLKovachy2015}.

Such manipulations have been experimentally implemented and allowed for monitoring the free evolution of a single-species BEC for 2\,s~\cite{van_zoest_bose-einstein_2010, PRLKovachy2015, Rudolph_2015}. In the case of a double-species BEC, we could consider applying the DKC atomic lens once, or using successive pulses, in order to control the expansion speed of the atomic clouds. Just as in the single-species case, the timing(s) and the duration(s) of the pulse(s) is of particular importance. The aim of this paper is to show that we can effectively collimate the two atomic ensembles such that they remain sufficiently compact after a free expansion of 10\,s. This time can be used to operate an efficient differential AI with the two species as proposed in the UFF test of Ref.~\cite{CQGAguilera2014, ASPALTSCHUL2015}. For such applications, magnetic disturbances of the clouds have to be avoided. To this end, the Feshbach magnetic field has to be ramped down after the last DKC pulse and the atomic ensembles transferred to magnetically insensitive states. One needs to verify that the density distribution of the two species keeps a suitable shape for the precision measurement despite the absence of the Feshbach field.

\section{Theoretical model}
\label{sec:theory}
\subsection{Mean-field equations}
\label{sec_3p1}

At zero temperature and within the mean-field approximation, the time evolution of a Bose-Einstein condensate is described by the time-dependent Gross-Pitaevskii equation (TD-GPE)~\cite{Pethick2002}
\begin{equation}
\label{Eq:TDGPE_general}
i\hbar\,\partial_t \Psi(\mathbf{r},t) =
\left[-\dfrac{\hbar^2}{2 m}\mathbf{\nabla}^2_{\mathbf{r}} + U(\mathbf{r},t) + Ng\,|\Psi(\mathbf{r},t)|^2\right] \Psi(\mathbf{r},t)\,,
\end{equation}
where $\Psi(\mathbf{r},t)$ denotes the wave function of the BEC, $m$ is the atomic mass, $U(\mathbf{r},t)$ the time dependent external potential, $N$ the number of particles in the BEC and $g$ the strength of the atom-atom interaction related at ultra-low temperature to the s-wave scattering length of the atomic species, $a$, by the relation $g = 4 \pi \hbar^2 a / m$. The wave function is normalized to one. In the initial potential, at $t=0$, the stationary solution is given by
\begin{equation}
\Psi(\mathbf{r},t) = \Psi(\mathbf{r},0)\,\exp\big[-i\,\mu\,t/\hbar\,\big]\,,
\end{equation}
where $\mu$ is the chemical potential of the quantum gas and the term $\exp[-i\mu t/\hbar]$ is a global phase. The time-independent GPE reads
\begin{equation}
\label{Eq_TIGPE}
\mu\,\Psi(\mathbf{r},0) =
\left[-\dfrac{\hbar^2}{2m}\mathbf{\nabla}^2_{\mathbf{r}} + U(\mathbf{r},0) + N g\,|\Psi(\mathbf{r},0)|^2 \right] \Psi(\mathbf{r},0)\,.
\end{equation}

\subsection{Single-species scaling approach}
\label{sec_3p3}

In the case of large atom numbers, the kinetic energy term of Eq.\,(\ref{Eq_TIGPE}) is much smaller than the interaction energy. In this so-called Thomas-Fermi (TF) limit, it is possible to express the probability density of the BEC as
\begin{equation}
\label{Eq_TF_wavefunction}
\rho^S(\mathbf{r},0) = N\,\big|\Psi^{\mathrm{TF}}(\mathbf{r},0)\big|^2 = 
\big[\mu-U(\mathbf{r},0)\big]/g\,,
\end{equation}
for $U(\mathbf{r},0) \leqslant \mu$ and $\rho^S(\mathbf{r},0) = 0$ otherwise. The exponent $S$ stands for the case of a single-species problem and will help to distinguished later the case of a double-species BEC. The chemical potential $\mu$ can be found by the normalization condition. For a harmonic trap this yields
\begin{equation}
\mu = \dfrac{\hbar\omega_0}{2}\left(\dfrac{15Na}{a_{\mathrm{osc}}}\right)^{2/5},
\end{equation}
where $a_{\mathrm{osc}}=\sqrt{\hbar/m\omega_0}$ is the average quantum-mechanical length scale of the 3D harmonic oscillator and where $\omega_0=\sqrt[3]{\omega_x(0)\,\omega_y(0)\,\omega_z(0)}$ is the geometric mean of the three oscillator frequencies in Cartesian coordinates~\cite{Pethick2002}. The size of the BEC along the directions $x$, $y$ and $z$ is characterized by the TF radii $R_x^{\,0}$, $R_y^{\,0}$ and $R_z^{\,0}$ given by
\begin{equation}
\label{eq_TF_radii}
R_\alpha^{\,0}=a_{\mathrm{osc}}\left(\dfrac{\omega_0}{\omega_\alpha(0)}\right)\left(\dfrac{15Na}{a_{\mathrm{osc}}}\right)^{\!\!1/5},
\end{equation}
with $\alpha \in \{x,y,z\}$.

In the TF approximation the parabolic shape of the density given in Eq.\,(\ref{Eq_TF_wavefunction}) remains unaltered when the frequencies of the harmonic trap vary and the cloud experiences a simple dilatation or a compression, which can be described by three scaling coefficients, $\lambda^S_{\alpha}(t)$, again with $\alpha \in \{x,y,z\}$. The size evolution of the cloud is then given by
\begin{equation}
\label{eq:TF_radii_scaling}
R_\alpha(t) = \lambda^S_\alpha(t)\;R_\alpha^{\,0}\,,
\end{equation}
and the evolution of the probability density is written as
\begin{equation}
\rho^S(\mathbf{r},t)  = \dfrac{\rho^S\left(\mathbf{r}\,',0\right)}{\lambda^S_x(t) \lambda^S_y(t) \lambda^S_z(t)}\;.
\end{equation}
In this expression the coordinates $\mathbf{r}$ and $\mathbf{r}\,'$ are defined as $\mathbf{r} = x\,\mathbf{u}_x + y\,\mathbf{u}_y + z\,\mathbf{u}_z$ and $\mathbf{r}\,' = (x/\lambda^S_{x}(t))\,\mathbf{u}_x+(y/\lambda^S_{y}(t))\,\mathbf{u}_y +(z/\lambda^S_{z}(t))\,\mathbf{u}_z$. Newton's law applied to the cloud dynamics yields
\begin{equation}
\ddot{\lambda}^S_\alpha(t)+\omega^2_\alpha(t)\lambda^S_\alpha(t) =
\dfrac{\omega^2_\alpha(0)}{\lambda^S_\alpha(t)\,\lambda^S_x(t)\lambda^S_y(t)\lambda^S_z(t)}\;,
\label{Scaling_law}
\end{equation}
where $\omega_{\alpha}(t)$ is the time-dependent trapping frequency in the direction $\alpha \in \{x,y,z\}$ ~\cite{castin_bose-einstein_1996, PhysRevA.55.R18}. The right hand side of Eq.\,(\ref{Scaling_law}) describes the coupling of the three directions through the mean field term of the GPE. Knowing the parabolic shape of the wave function, the three typical sizes $R_\alpha(t)$ can be related to the three standard deviations $\Delta\alpha(t)$ of the BEC density. This relation is $\Delta\alpha(t) = R_\alpha(t)\,/\,\sqrt{7}$. Numerically, we also evaluate these three widths $\Delta x(t)$, $\Delta y(t)$ and $\Delta z(t)$ from the solution of the time-dependent Gross-Pitaevskii equation\,(\ref{Eq:TDGPE_general}).

\subsection{Coupled mean-field equations}

In the last section the model used to describe single BEC dynamics has been presented. A similar set of coupled equations can be derived to study the dynamics of an interacting mixture of degenerate gases. In the case of a two-component BEC, within the mean field approximation, the dynamics is described by the time-dependent coupled Gross-Pitaevskii equations (TD-CGPE)
\begin{subequations}
\label{eq_TDCGPE}
\begin{align}
\label{eq_TDCGPE_a}
i \hbar \partial_t \Psi_1(\mathbf{r},t) & = \Big[-\dfrac{\hbar^2\mathbf{\nabla}^2_{\mathbf{r}}}{2 m_1} +U_1(\mathbf{r},t)+N_1 g_{11}|\Psi_1(\mathbf{r},t)|^2+N_2 g_{12}|\Psi_2(\mathbf{r},t)|^2\Big]\Psi_1(\mathbf{r},t)\\
\label{eq_TDCGPE_b}
i \hbar \partial_t \Psi_2(\mathbf{r},t) & = \Big[-\dfrac{\hbar^2\mathbf{\nabla}^2_{\mathbf{r}}}{2 m_2} +U_2(\mathbf{r},t)+N_2 g_{22}|\Psi_2(\mathbf{r},t)|^2+N_1 g_{12}|\Psi_1(\mathbf{r},t)|^2\Big]\Psi_2(\mathbf{r},t)
\end{align}
\end{subequations}
where $\Psi_i(\mathbf{r},t)$ with $i \in \{ 1,2\}$ denotes the wave function of the species number $i$. The constants $g_{ij}$ are related to the respective scattering lengths, $a_{11}$, $a_{12}$ and $a_{22}$ by the relation $g_{ij}=2\pi\hbar^2a_{ij}/m_{ij}$, with $m_{ij}$ being the reduced mass $m_i m_j/(m_i+m_j)$. $N_i$ and $U_i(\mathbf{r},t)$ are the number of atoms and the external potential of the species $i$, respectively. In the following we consider the two condensates to be confined in external harmonic traps of frequencies $\omega_{i,\alpha}(t)$. The last terms of Eqs.\,(\ref{eq_TDCGPE_a}) and\,(\ref{eq_TDCGPE_b}) describe the coupling between the two components. Both wave functions are normalized to 1. For large atom numbers and within the TF approximation, the time-independent coupled Gross-Pitaevskii equations read
\begin{subequations}
\label{eq_TICGPE_TF}
\begin{align}
\label{eq_TICGPE_TF_a}
\mu_1 & = U_1(\mathbf{r},0)+N_1 g_{11}|\Psi_{1}^{\mathrm{TF}}(\mathbf{r},0)|^2+N_2 g_{12}|\Psi_{2}^{\mathrm{TF}}(\mathbf{r},0)|^2\,,\\
\mu_2 & = U_2(\mathbf{r},0)+N_2 g_{22}|\Psi_{2}^{\mathrm{TF}}(\mathbf{r},0)|^2+N_1 g_{12}|\Psi_{1}^{\mathrm{TF}}(\mathbf{r},0)|^2\,.
\label{eq_TICGPE_TF_b}
\end{align}
\end{subequations}

The nature of the solution of Eq.\,(\ref{eq_TICGPE_TF}) is determined by the competition between the intra- and inter-species interactions. If the single-species interaction dominates ($g_{11}g_{22} > g_{12}^2$ for a uniform gas) the energy is minimized when the two species occupy the entire accessible volume. In this case the two BEC wave functions overlap and we are in the miscible regime~\cite{PhysRevLett.81.5718, PhysRevA.58.4836, JPB_Trippenbach_2000}. Conversely, if the inter-species interaction dominates ($g_{12}^2 > g_{11}g_{22}$ for a uniform gas) the energy of the system is minimized when the two BEC wave functions are spatially separated. This is the immiscible regime~\cite{PhysRevLett.81.5718, PhysRevA.58.4836, JPB_Trippenbach_2000}. Having precision interferometry applications as a motivation, the present study is realized in the miscible regime and we consider that the three following conditions are fulfilled: $g_{11}>0$, $g_{22}>0$ and $G^2 =g_{11}\,g_{22}-g_{12}^2>0$. In this case the density distribution of the two interacting condensates can be approximated in a similar fashion as in the case of a single component condensate.  Therefore, in the overlap region the density distributions of the two species are given by
\begin{subequations}
\label{Eq_TF_wavefunction_mixture}
\begin{align}
\label{Eq_TF_wavefunction_mixture1}
\rho_1^D(\mathbf{r},0) = N_1\,|\Psi_{1}^{\mathrm{TF}}(\mathbf{r},0)|^2 & = 
\dfrac{g_{22}}{G^2}\,[\mu_1-U_1(\mathbf{r},0)]-\dfrac{g_{12}}{G^2}\,[\mu_2-U_2(\mathbf{r},0)]\\
\label{Eq_TF_wavefunction_mixture2}
\rho_2^D(\mathbf{r},0) = N_2\,|\Psi_{2}^{\mathrm{TF}}(\mathbf{r},0)|^2 & = 
\dfrac{g_{11}}{G^2}\,[\mu_2-U_2(\mathbf{r},0)]-\dfrac{g_{12}}{G^2}\,[\mu_1-U_1(\mathbf{r},0)]
\end{align}
\end{subequations}
for $g_{22}[\mu_1-U_1(\mathbf{r},0)]>g_{12}[\mu_2-U_2(\mathbf{r},0)]$ and $g_{11}[\mu_2-U_2(\mathbf{r},0)]>g_{12}[\mu_1-U_1(\mathbf{r},0)]$. These two conditions define the region of co-existence of the two BECs, where Eqs. (\ref{Eq_TF_wavefunction_mixture1}) and (\ref{Eq_TF_wavefunction_mixture2}) hold. In other regions, the densities are given by the single component of Eq. (\ref{Eq_TF_wavefunction}).

\subsection{Dual-species scaling approach}

\begin{figure}[t!]
\includegraphics[width=1.0\columnwidth]{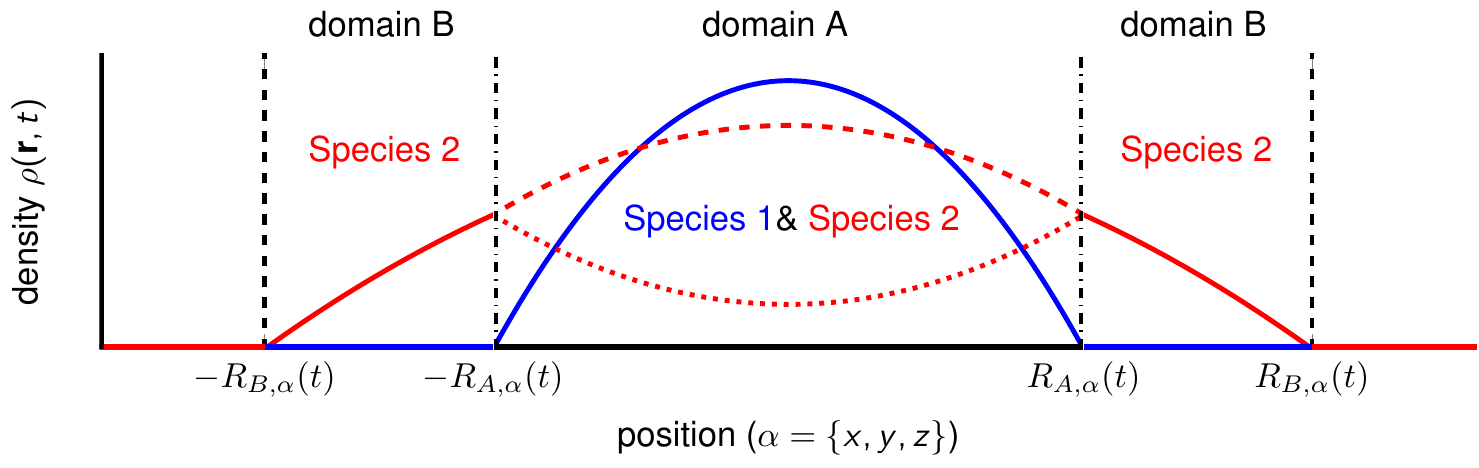}
\caption{Representation of the different domains and density configurations of the two species. In the central domain (domain A) both species are present and the curvature of the second species (red) depends on the interspecies and intraspecies s-wave scattering length\,\cite{JPB_Trippenbach_2000,CorgierPHDthesis}. In the outer domain (domain B) only the second species is present.}
\label{fig_0}
\end{figure}

We present now a generalization of the scaling theory introduced in Refs.~\cite{castin_bose-einstein_1996, PhysRevA.55.R18} in order to account for the mutual interactions between the two species in the region of overlap. As highlighted in Fig.\,\ref{fig_0} we consider two distinct domains: In the central one corresponding to domain\;A, the two species are present. In the second domain called domain B, which surrounds the inner domain A, only one of the two species is present, namely species number 2. This naturally leads us to define six scaling factors $\lambda^D_{i,\alpha}(t)$ for the domain A, with $i \in \{1,2\}$ and $\alpha \in \{x,y,z\}$, and three scaling factors $\lambda^S_{2,\alpha}(t)$ for the domain B. The exponent $D$ denotes the presence of the two species in domain A and the exponent $S$ is for the outer domain B with a single species. Since domain B is characterized by the presence of a single species, the evolution dynamics of the scaling factors $\lambda^S_{2,\alpha}(t)$ is governed by Eq. (\ref{Scaling_law}).

In analogy with Eqs. (\ref{eq_TF_radii}) and (\ref{eq:TF_radii_scaling}), we denote respectively by $R_{A,\alpha}(t)$ and $R_{B,\alpha}(t)$ the outer limits of the two domains in the direction $\alpha$ at time $t$ according to
\begin{subequations}
\label{eq_Radii_Sph_R_t}
\begin{align}
\label{eq_Radii_Sph_R_t_a}
R_{A,\alpha}(t) & = \lambda_{1,\alpha}^D(t)\;R_{A,\alpha}^{\,0}\\
\label{eq_Radii_Sph_R_t_b}
R_{B,\alpha}(t) & = \lambda_{2,\alpha}^S(t)\;R_{B,\alpha}^{\,0}
\end{align}
\end{subequations}
where the initial sizes $R_{A,\alpha}^{\,0}$ of domain A are defined by the cancellation of the density distribution of the first species: $\rho_1^D(\mathbf{r},0)=0$ for $\mathbf{r} \in \big\{R_{A,x}^{\,0}\,\mathbf{u}_x\,; R_{A,y}^{\,0}\,\mathbf{u}_y\,; R_{A,z}^{\,0}\,\mathbf{u}_z\big\}$. Similarly, the initial sizes $R_{B,\alpha}^{\,0}$ of domain B are defined by the cancellation of the density distribution of the second species: $\rho_2^S(\mathbf{r},0)=0$ for $\mathbf{r} \in \big\{R_{B,x}^{\,0}\,\mathbf{u}_x\,; R_{B,y}^{\,0}\,\mathbf{u}_y\,; R_{B,z}^{\,0}\,\mathbf{u}_z\big\}$.

Following the different steps of calculation described in Refs.~\cite{castin_bose-einstein_1996, PhysRevA.55.R18} for a single component BEC, we obtain in the case of a double species BEC six coupled first order differential equations~\cite{CorgierPHDthesis}
\begin{align}
\label{Eq_Scaling_Dual}
\ddot{\lambda}_{\;i,\alpha}^D(t) = & -\omega_{i,\alpha}^2(t)\,\lambda_{i,\alpha}^D(t) \nonumber \\
& + \dfrac{g_{ii}}{\lambda_{i,x}^D(t)\,\lambda_{i,y}^D(t)\,\lambda_{i,z}^D(t)} \left(\dfrac{g_{i'i'}\,m_i\,\omega_{i,\alpha}^2(0)\,-\,g_{12}\,m_{i'}\,\omega_{i',\alpha}^2(0)}{m_i\,G^2\lambda_{i,\alpha}^D(t)}\right) \nonumber \\
& + \dfrac{g_{12}}{\lambda_{i',x}^D(t)\,\lambda_{i',y}^D(t)\,\lambda_{i',z}^D(t)} \left(\dfrac{g_{ii}\,m_{i'}\,\omega_{i',\alpha}^2(0)\,-\,g_{12}\,m_i\,\omega_{i,\alpha}^2(0)}
{m_i\,G^2\lambda_{i',\alpha}^D(t)}\right)\dfrac{\lambda_{i,\alpha}^D(t)}{\lambda_{i',\alpha}^D(t)}
\end{align}
describing the time evolution of the six scaling factors $\lambda^D_{i,\alpha}(t)$ in domain A. In this equation $i'=2$ when $i=1$ and $i'=1$ when $i=2$. It is easily seen that if the two BEC components do not interact (case where $g_{12}=0$), Eq.\,(\ref{Eq_Scaling_Dual}) reduces to a set of single component equations identical to Eq.\,(\ref{Scaling_law}). By solving these six coupled first order differential equations (\ref{Eq_Scaling_Dual}) we are able to follow the size dynamics of the two-component condensate without having to solve the coupled Gross-Pitaevskii equations (\ref{eq_TDCGPE}). As we will show in the next section, this turns out to be a reliable approximation which provides a computationally efficient predictions of the two-species condensate expansion or compression dynamics as long as the number of domains (\emph{i.e.} 2 domains) is conserved.

The evolution of the atomic densities of the two species in domain A is given by
\begin{subequations}
\label{eq_SA_density_overlap_t}
\begin{align}
\label{eq_SA_density_overlap_t_a}
\rho_1^D(\mathbf{r},t) & = \dfrac{\rho_1^D(\mathbf{r}\,',0)}{\lambda^D_{1,x}(t) \lambda^D_{1,y}(t) \lambda^D_{1,z}(t)}
\quad\mathrm{with}\quad
\mathbf{r}\,' = \dfrac{x\,\mathbf{u}_x}{\lambda^D_{1,x}(t)} +  \dfrac{y\,\mathbf{u}_y}{\lambda^D_{1,y}(t)} +  \dfrac{z\,\mathbf{u}_z}{\lambda^D_{1,z}(t)}\\
\label{eq_SA_density_overlap_t_b}
\rho_2^D(\mathbf{r},t) & = \dfrac{\rho_2^D(\mathbf{r}\,'',0)}{\lambda^D_{2,x}(t) \lambda^D_{2,y}(t) \lambda^D_{2,z}(t)}
\quad\mathrm{with}\quad
\mathbf{r}\,'' = \dfrac{x\,\mathbf{u}_x}{\lambda^D_{2,x}(t)} +  \dfrac{y\,\mathbf{u}_y}{\lambda^D_{2,y}(t)} +  \dfrac{z\,\mathbf{u}_z}{\lambda^D_{2,z}(t)}
\end{align}
\end{subequations}
Here $\mathbf{r}\,'$ and $\mathbf{r}\,''$ denote re-scaled coordinates compared to $\mathbf{r} = x\,\mathbf{u}_x + y\,\mathbf{u}_y + z\,\mathbf{u}_z$. In the outer domain B we obtain
\begin{subequations}
\label{eq_SA_density_no_overlap_t}
\begin{align}
\label{eq_SA_density_no_overlap_t_a}
\rho_1^S(\mathbf{r},t) & = 0\\
\label{eq_SA_density_no_overlap_t_b}
\rho_2^S(\mathbf{r},t) & =\dfrac{\rho_2^S(\mathbf{r}\,''',0)}{\lambda^S_{2,x}(t) \lambda^S_{2,y}(t) \lambda^S_{2,z}(t)}
\quad\mathrm{with}\quad
\mathbf{r}\,''' = \dfrac{x\,\mathbf{u}_x}{\lambda^S_{2,x}(t)} +  \dfrac{y\,\mathbf{u}_y}{\lambda^S_{2,y}(t)} +  \dfrac{z\,\mathbf{u}_z}{\lambda^S_{2,z}(t)}\,.
\end{align}
\end{subequations}

\subsection{Delta-kick collimation}

To largely reduce the expansion rate of cold atomic samples, the delta-kick collimation (DKC) technique~\cite{chu_proposal_1986, ammann_delta_1997,NJPCorgier2018} is commonly applied. It consists in re-trapping a freely expanding cloud of atoms for a brief duration in order to align its phase-space density distribution along the position coordinate axis, therefore minimizing its momentum distribution width in preparation for a further expansion. This is in analogy with the collimation effect of a lens in optics and DKC is often referred to as an atomic lens. It is worth noticing that the phase-space density of lensed ensemble is conserved which does not qualify this process to be a cooling in the strict statistical physics sense. This method was successfully implemented and led to record-long observation times of several seconds~\cite{PRLMuntinga2013,PhysRevLett.114.143004, Rudolph-thesis}. The DKC effect is accounted for in the dynamics by simply considering the time-dependent trap frequencies defined as follows: $\omega_{\,i,\alpha}(t) = \omega_{\,i,\alpha}(0)$ if $t^i_\mathrm{DKC}\leqslant t \leqslant t^f_\mathrm{DKC}$ and $\omega_{\,i,\alpha}(t) = 0$ during the free expansion. Here, $t^{i(f)}_\text{DKC}$ is the starting (final) time of application of the optical lens.

\subsection{Feshbach magnetic field}

Reference~\cite{Thalhammer_2008} reports the discovery of two Feshbach resonances around 35\,G and 79\,G in a mixture of $^{41}$K and $^{87}$Rb, where the mutual interaction, with magnitude $a_{12}$, is magnetic-field-dependant. This instrumental feature will be used in the following in order to switch-off the K-Rb interaction at short times to enhance miscibility. We will thus consider a sequence where $a_{12}(t) = 0$ for $t \leqslant t_\mathrm{F}$ and $a_{12}(t) = 163\,a_0$ for $t > t_\mathrm{F}$. Indeed, by switching-off the external magnetic field $B_0$ at time $t_\mathrm{F}$, the inter-species interaction is naturally at the latter value. The s-wave scattering lengths of Rubidium and Potassium are constant in the vicinity of the abovementioned Feshbach resonances and respectively equal to $a_{\mathrm{Rb}}=a_{11}=99\,a_0$ and $a_{\mathrm{K}}=a_{22}=60\,a_0$~\cite{PRLFerrari2002, PRAFerlaino2006, Thalhammer_2008, dataRb}.

\subsection{Numerical considerations}

Two methods are used to describe the ground state or dynamics of the condensates. The solutions of the TD-CGPE are propagated using the split-operator method reported in~\cite{feit_solution_1982} by means of fast Fourier transforms. To find the ground state of the mixture, the propagation is carried out in imaginary time, so as to let the solution relax to the ground state following the approach of Ref.\,\cite{JCPLehtovaara2007}.
This solution is then used as the initial state of the real-time propagation. Solutions to the scaling equations are obtained using a fourth order Runge-Kutta integrator. More details on the numerical algorithm can be found in~\cite{CorgierPHDthesis}.

\section{Engineered free expansion of a binary mixture}
\label{sec:results}

The binary mixture described theoretically in the Sec.\,\ref{sec:theory} is designed as the input of an atomic interferometer (AI) dedicated to high-precision measurements similar to the UFF test of reference~\cite{ASPALTSCHUL2015}. The use of a large magnetic field during operation of the AI is not possible due to several systematic effects that appear in relation with Zeeman shifts. This field is however essential for the preparation of the mixture to overcome the problem of immiscibility and shape deformations of the density distribution of the two species after release, during the free-expansion time of the interferometry sequence. Indeed, the deformation of the distribution can lead to detrimental wave-front aberrations~\cite{Louchet_Chauvet_2011} such as the appearance of inhomogeneous phases imprinted by the pulses of the interferometry sequence. Moreover, for state-of-the-art precision measurements, a long-time atom interferometer is required and one would therefore benefit from slow kinetic expansion rates of the two atomic ensembles as delivered by DKC. Long-time atom interferometers are nowadays accessible on Earth. In the case of a Mach-Zehnder-type atom interferometer, the total interferometry time can be of the order of 2\,s but its successful operation requires the control of the environment over a 10-meter long experiment~\cite{allen1987optical, Asenbaum2020arxiv}. In micro-gravity environments~\cite{van_zoest_bose-einstein_2010, Rudolph_2015} or in space~\cite{NatureBecker2018, CALwebsite, Battelier2019}, longer interrogation times of about 10 seconds are available and would be considered in what follows.

\subsection{Isotropic trap}

To simplify the description of the dual-species theoretical treatment of its dynamics, we choose the external trap to be harmonic and isotropic as it could be realised by crossed optical traps for example~\cite{Li2019}. For the atom number in species $i$ ($i=1$ for Rb and $i=2$ for K) we denote by $\omega_i(t)$ the associated frequency, such that $U_i(r,0)=m_i\,\omega_i^2(0)\,r^2/2$. In spherical coordinates one can write the wave function as a product of radial and angular parts such as $\Psi_i(\mathbf{r},t) = \chi_i(r,t) \cdot Y_{\ell,m}(\theta,\phi)\,/\,r$, where, in the particular case of a pure spherical trap, $\ell=m=0$. This transformation leads to solve a simplified one dimensional radial TD-CGPE
\begin{subequations}
\label{eq_TDCGPE_radial}
\begin{align}
\label{eq_TDCGPE_a_radial}
i\hbar\partial_t \chi_1(r,t) & = \left[-\dfrac{\hbar^2\partial_{rr}}{2m_1} + U_1(r,t)
+\dfrac{N_1 g_{11}}{4\pi r^2}|\chi_1(r,t)|^2
+\dfrac{N_2 g_{12}}{4\pi r^2}|\chi_2(r,t)|^2\right]\chi_1(r,t)\\
\label{eq_TDCGPE_b_radial}
i\hbar\partial_t \chi_2(r,t) & = \left[-\dfrac{\hbar^2\partial_{rr}}{2m_2} + U_2(r,t)
+\dfrac{N_2 g_{22}}{4\pi r^2}|\chi_2(r,t)|^2
+\dfrac{N_1 g_{12}}{4\pi r^2}|\chi_1(r,t)|^2\right]\chi_2(r,t)
\end{align}
\end{subequations}
with the normalization conditions
\begin{equation}
\int_{0}^{\infty} |\chi_i(r,t)|^2\,dr = 1 \quad \mathrm{for} \quad i=1,2\,.
\end{equation}
Using the Thomas-Fermi approximation, the initial density distributions $\rho_1^D(r,0)$ and $\rho_2^D(r,0)$ of species 1 and 2 in domain A are still given by Eqs.\,(\ref{Eq_TF_wavefunction_mixture1}) and (\ref{Eq_TF_wavefunction_mixture2}), while the initial density distribution $\rho_2^S(r,0)$ of species 2 in domain B is given by the single species expression (\ref{Eq_TF_wavefunction}). The only difference is that the vector $\mathbf{r}$ in Eqs.\,(\ref{Eq_TF_wavefunction}), (\ref{Eq_TF_wavefunction_mixture1}) and (\ref{Eq_TF_wavefunction_mixture2}) is replaced by the radial coordinate $r$. In this case of an isotropic harmonic trap, the initial Thomas-Fermi radii are defined by
\begin{subequations}
\label{eq_Radii_Sph_R_0}
\begin{align}
\label{eq_Radii_Sph_R_0_a}
R_A^{\,0}&=\left[\dfrac{2\,(g_{22}\mu_1+g_{12}\mu_2)}{g_{22}m_1\omega_1^2(0)-g_{12}m_2\omega_2^2(0)}\right]^{1/2}\\
\label{eq_Radii_Sph_R_0_b}
R_B^{\,0}&=\left[\dfrac{2\,\mu_2}{m_2\omega_2^2(0)}\right]^{1/2}
\end{align}
\end{subequations}
for domains A and B. In the same way as for single species, the chemical potential is found thanks to the normalization conditions
leading to
\begin{subequations}
\begin{align}
\mu_1 & =\dfrac{g_{12}}{g_{22}}\mu_2+\left(\dfrac{15}{8\pi}\dfrac{N_1G^2}{g_{22}}\right)^{\!2/5}\left(\dfrac{m_1\omega_1^2(0)}{2} - \dfrac{g_{12}m_2\omega_2^2(0)}{2g_{22}}\right)^{\!3/5},\\[0.3cm]
\mu_2 & =\left(\dfrac{15}{8\pi}(N_2 g_{22} + N_1 g_{12})\right)^{\!2/5}\left(\dfrac{m_2 \omega_2^2(0)}{2}\right)^{\!3/5}.
\end{align}
\end{subequations}

The scaling equations (\ref{Eq_Scaling_Dual}) describing the size dynamics of the binary mixture simplify into
\begin{align}
\label{Eq_Scaling_Dual_isotrop}
\ddot{\lambda}_{\;i}^D(t) + \omega_{i}^2(t)\,\lambda_{i}^D(t) & = 
\dfrac{g_{ii}}{\big[\lambda_{i}^D(t)\big]^4} \left(\dfrac{g_{i'i'}\,m_i\,\omega_{i}^2(0)\,-\,g_{12}\,m_{i'}\,\omega_{i'}^2(0)}{m_i\,G^2}\right) \nonumber \\
& + \;\dfrac{g_{12}\,\lambda_{i}^D(t)}{\big[\lambda_{i'}^D(t)\big]^5} \left(\dfrac{g_{ii}\,m_{i'}\,\omega_{i'}^2(0)\,-\,g_{12}\,m_i\,\omega_{i}^2(0)}
{m_i\,G^2}\right),
\end{align}
where, in the following, we treat the case $m_i\,\omega_{i}^2(0)=m_{i'}\,\omega_{i'}^2(0)$ leading to
\begin{equation}
\label{Eq_Scaling_Dual_isotrop_simple}
\ddot{\lambda}_{\;i}^D(t) + \omega_{i}^2(t)\,\lambda_{i}^D(t) =
\left[\dfrac{g_{ii}(g_{i'i'}-g_{12})}{\big[\lambda_{i}^D(t)\big]^4} +
\dfrac{g_{12}(g_{ii}-\,g_{12})\lambda_{i}^D(t)}{\big[\lambda_{i'}^D(t)\big]^5}\right]
\dfrac{\omega_{i}^2(0)}{G^2}\,.
\end{equation}
Finally, the sizes of the two domains are obtained from the calculation of the Thomas-Fermi radii $R_A(t)=\lambda_{1}^D(t)\;R_A^{\,0}$ and $R_B(t)=\lambda_{2}^S(t)\;R_B^{\,0}$. The density distributions $\rho_1^D(r,t)$ and $\rho_2^D(r,t)$ of species 1 and 2 in domain A are given by Eqs.\,(\ref{eq_SA_density_overlap_t_a}) and (\ref{eq_SA_density_overlap_t_b}), while the density distribution $\rho_2^S(r,t)$ of species 2 in domain B is given by Eq.\,(\ref{eq_SA_density_no_overlap_t_b}).

To compare to the results of this scaling approach specifically designed for binary mixtures with the TD-CGPE (\ref{eq_TDCGPE}), it is convenient to define the characteristic standard deviations $\Delta r_1(t)$ and $\Delta r_2(t)$ of the density distributions of species 1 and 2. Indeed, these characteristic sizes can be calculated either from the densities obtained with the TD-CGPE or from the densities obtained with the generalized scaling approach.

\subsection{Initial state}

Before looking at the dynamics of the expanding source, we first study the initial stationary binary mixture confined in a harmonic and isotropic trap. The ground states of this quantum mixture are not trivial and deserve a careful description, especially when the miscibility of the two quantum fluids comes into play~\cite{PhysRevLett.81.5718, PhysRevA.58.4836, JPB_Trippenbach_2000}. An immiscible mixture is not a suitable source for a high-precision interferometer since an offset between the center-of-mass of the two gases couples to gravity or magnetic field gradients leading to important systematic errors~\cite{CQGAguilera2014}. In our case we consider the two-component super-fluid to be in the miscible regime characterized by the same center-of-mass and by the same domain of existence.

\begin{figure}[t!]
\includegraphics[width=\textwidth]{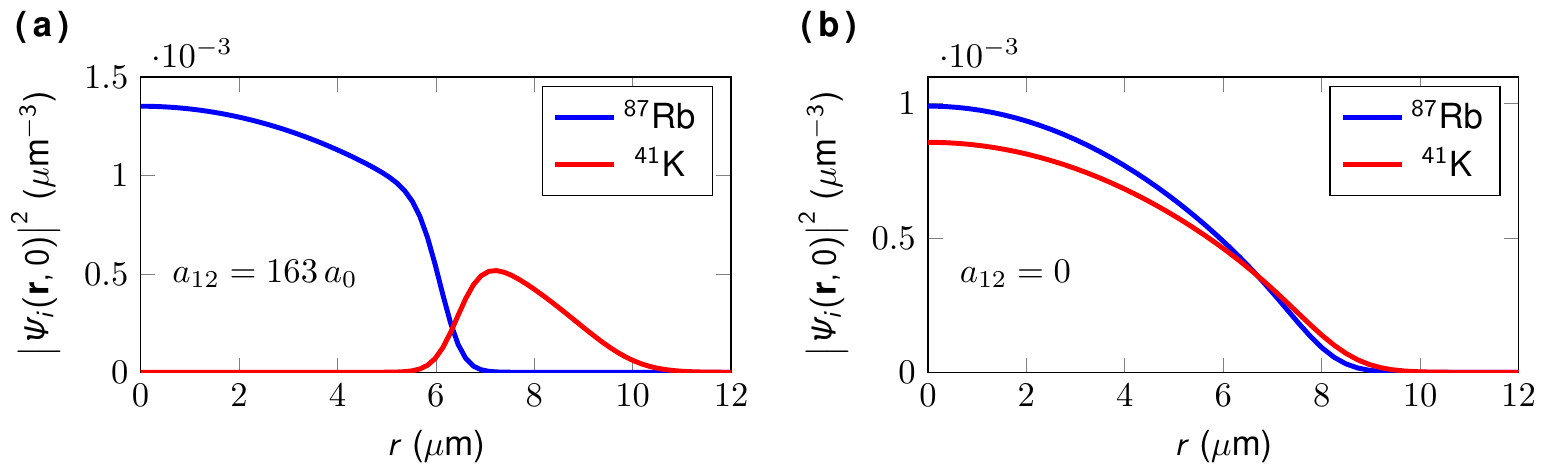}
\caption{Ground-state of a binary mixture of $^{87}$Rb and $^{41}$K. In the left panel (a) $a_{12}=163\,a_0$ and the two BECs repel each other. In the right panel (b) the inter-species scattering length is tuned to zero and the two BECs do not interact. The calculations have been done with $\omega_1(0)\equiv\omega_{\mathrm{Rb}} = 2\pi \times 50$\,Hz and $\omega_2(0)\equiv\omega_{\mathrm{K}} = 2\pi \times 73$\,Hz, with $10^5$ atoms in each BEC. The blue and red colors are for the Rb and K species, respectively.}
\label{fig_GS}
\end{figure}

In Fig.\,\ref{fig_GS}, we show how the inter-species interaction length $a_{12}$ impacts the ground state density distribution obtained by solving the coupled Gross–Pitaevskii equations. Here the mixture is created in a trap with mean frequencies $\omega_1(0) \equiv \omega_{\mathrm{Rb}} = 2\pi \times 50$\,Hz and $\omega_2(0) \equiv \omega_{\mathrm{K}} = (m_{\mathrm{Rb}}/m_{\mathrm{K}})^{1/2} \omega_{\mathrm{Rb}} \simeq 2\pi \times 73$\,Hz for Rb and K, respectively. In the left panel (a) the Feshbach field is turned off and $a_{12}=163\,a_0$. The contact interaction energy is then dominated by the repulsion between Rb and K atoms, and the two BECs repel each other. As a consequence, the two BECs do not overlap: the Rb-BEC is located at the center of the trap, surrounded by the K-BEC, as seen in Fig.\,\ref{fig_GS}(a). This state is fragile against external perturbations and can lead in the non-ideal experimental environment to an immiscible, asymmetric state where the two gases are located side-by-side. 

In the right panel (b) of the same figure, the Feshbach magnetic field is chosen such that $a_{12}=0$. In this case the two BECs do not interact with each other and they maximally overlap since they essentially share the same domain of existence. As we will see later on, if this cold atomic mixture was released with the Feshbach magnetic field turned off, both spatial distributions would be highly modified due to strong inter-species repulsion, quickly leading to a spatial separation of the two species similar to the one seen in Fig.\,\ref{fig_GS}(a).

\begin{figure}[t!]
\includegraphics[width=\textwidth]{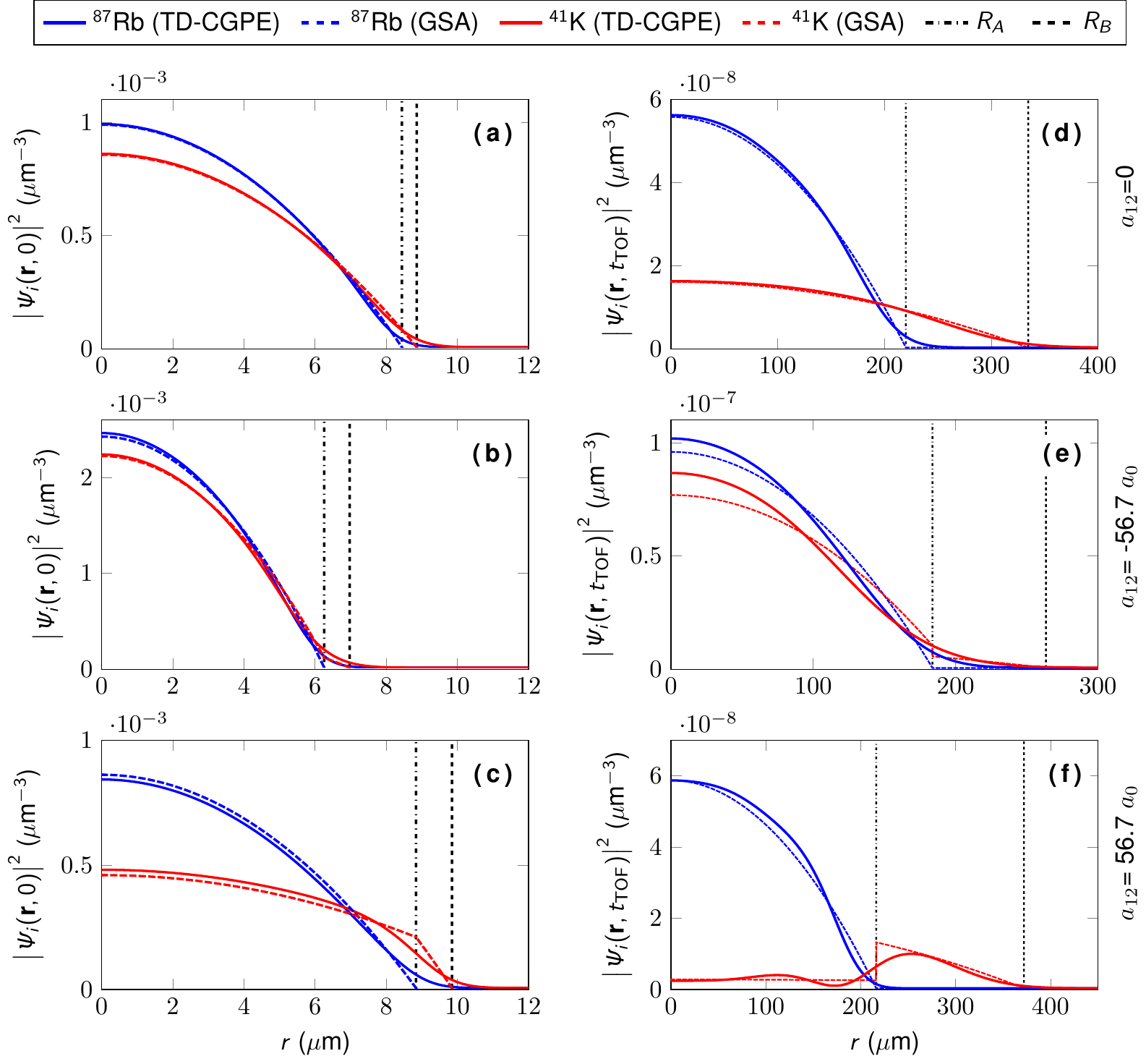}
\caption{Representation of the density distribution of the BEC mixture in different situations. The left panel is the initial density at $t=0$. The right panel is after a Time-of-Flight (TOF) $t_{\mathrm{TOF}}=100$\,ms. The blue and red colors are for Rb and K. The straight and dashed lines show the results of the calculations with the TD-CGPE approach and with the generalized scaling approach (GSA), respectively. The inter-species scattering length is $a_{12}=0$ in the upper panels (a) and (d), $a_{12}=-56.7\,a_0$ in middle panels (b) and (e) and $a_{12}=+56.7\,a_0$ in the lower panels (c) and (f). The vertical dash-dotted and dashed lines mark the expected sizes of the two domains, $R_A(t)$ and $R_B(t)$.}
\label{fig_GS_100ms_TOF_NE5_V3}
\end{figure}

\subsection{Expansion dynamics}

In this section we focus on the case where the two BECs overlap at all times. This regime is defined by the criterion $G^2=g_{11}\,g_{22}-g_{12}^2>0$, equivalent in our Rb-K case to $a_{12} < 72\,a_0$. This criterion was derived for uniform gases in homogeneous traps and considers only the different mean-field interactions present in the system~\cite{PhysRevLett.81.5718, PhysRevA.58.4836, JPB_Trippenbach_2000}. Since most of the experiments operate in harmonic traps, the density of the atomic clouds is far from having a uniform profile. In this case, the miscibility is highly dependent on the number of atoms as well. We nevertheless use this criterion as a rough reference in the following. As in Fig.\,\ref{fig_GS}, we consider the case of a  mixture created initially in a trap with mean frequencies $\omega_1(0) \equiv \omega_{\mathrm{Rb}} = 2\pi \times 50$\,Hz and $\omega_2(0) \equiv \omega_{\mathrm{K}} = 2\pi \times 73$\,Hz. The left column of Fig.\,\ref{fig_GS_100ms_TOF_NE5_V3} shows the initial density distributions of the two species for an inter-species scattering length tuned from $a_{12}=0$ in panel (a) to $a_{12} = -56.7\,a_0 = -3$\,nm in panel (b) and to $a_{12} = +56.7\,a_0 = +3$\,nm in panel (c). In each panel the blue and red lines are for Rb and K, respectively. The solid and dashed lines show the solutions of the coupled Gross-Pitaevskii equations (\ref{eq_TDCGPE}) and of the generalized scaling approach (\ref{Eq_TF_wavefunction_mixture}), respectively. The vertical dash-dotted  and dotted lines mark the limits of domains A and B.

We now verify the accuracy of the generalized scaling approach in the case of a free expansion of the two condensates. To this end, we first calculate the expansion dynamics using the time-dependent coupled Gross-Pitaevskii equations and then compare to the generalized scaling equations for different ground state configurations shown on the left side of Fig.\;\ref{fig_GS_100ms_TOF_NE5_V3}. The right side of Fig.\;\ref{fig_GS_100ms_TOF_NE5_V3} shows the corresponding density distributions calculated with the TD-CGPE (solid lines) and with the generalized scaling approach (dashed lines) after a Time-of-Flight (TOF) $t_{\mathrm{TOF}}=100$\,ms in the case where the inter-species scattering length is tuned from $a_{12}=0$ in the upper panel (d) to $a_{12}=-56.7\,a_0$ in the middle panel (e) and to $a_{12}=+56.7\,a_0$ in the lower panel (f).

As expected, it can first be noticed that the generalized scaling approach is very accurate for $a_{12}=0$. This approach also provides a rather good and almost quantitative description of the two-species cloud expansion dynamics for the cases presented in panels (e) and (f) with $g_{11}g_{22}/g_{12}^2 \simeq 1.6$. However, when $a_{12} \neq 0$, the density distribution predicted for K by the generalized scaling approach becomes discontinuous at the boundary between the two domains. This comes from the fact that the spatial density of K is described by Eq.\,(\ref{eq_SA_density_overlap_t_b}) in the inner domain A and by Eq.\,(\ref{eq_SA_density_no_overlap_t_b}) in the outer domain B. The Rb species, which is not present in domain B does not show such a discontinuity. We see here that when $g_{12}^2 < g_{11}g_{22}$ the main features of the density distributions of the two species are caught by the generalized scaling approach but not their fine details such as the oscillation of the K density seen in domain A in Fig.\,\ref{fig_GS_100ms_TOF_NE5_V3}(f) for instance. Such an accuracy level is however sufficient to predict the expansion rates of the two components~\cite{CorgierPHDthesis}. In addition, solving the generalized scaling approach is numerically much more efficient than solving the time-dependent coupled Gross-Pitaevskii equations.

We also note that in the case of negative inter-species scattering length the expansion rates of the two clouds are reduced by the inter-species attraction. This feature can be qualitatively understood if one interprets the inter-species mean-field energy, \emph{i.e.} the last term of Eq.\,(\ref{eq_TDCGPE}), as a confining potential. Nevertheless, this result  has to be considered with caution since we only account for mean-field interactions in this study. Considering the first order Lee-Huang-Yang correction to the mean field approach\,\cite{PhysRev.106.1135}, a creation of quantum droplets in an attractive mixture has recently been predicted~\cite{PRL_Petrov_2015} and investigated in the case of $^{87}$Rb and $^{41}$K~\cite{PRErrico2019}, showing a stabilization of the mixture instead of a collapse.

It should also be emphasized that for $a_{12}\neq 0$, the generalized scaling approach does not conserve the total number of atoms. This feature is inherent to the model. Indeed, this model assumes that there are initially two separate, uncoupled domains, for which a separate scaling approach is performed. However, in reality, since the expansion dynamics is different for Rb and K, it can happen that a fraction of the atoms of a given species leaves one of the domains in favor of the other one. This phenomenon is naturally taken into account in the TD-CGPE approach but not in the generalized scaling approach which simply consists of associating two scaling parameters $\lambda_{\mathrm{Rb}}^A$ and $\lambda_{\mathrm{K}}^A$ to domain A and one scaling parameter $\lambda_{\mathrm{K}}^B$ to domain B. In the results presented in Fig.\,\ref{fig_GS_100ms_TOF_NE5_V3}, at the end of a TOF dynamics, we obtain $\lambda_{\mathrm{K}}^A > \lambda_{\mathrm{Rb}}^A$, meaning that the K cloud initially in domain A expands faster than the Rb cloud. Since domain A is defined as the domain shared by the two species, we can conclude that a fraction of the K atoms initially in domain A leaves this domain during the TOF dynamics. The number of K atoms in domain A therefore decreases with time. This population transfer is not accounted for by the model since it does not include any term coupling the two different domains. Similarly, the number of K atoms in domain B is not constant either. This is obviously one of the limitations of this model. This limitation may be usefully used as a measure of the model accuracy: In the three cases shown Fig.\,\ref{fig_GS_100ms_TOF_NE5_V3}, after 100\,ms of TOF  we obtain $\Delta N_\mathrm{K}\,/\,N_\mathrm{K} = 0$ when $a_{12}=0$ [panel\,(d)], $\Delta N_\mathrm{K}\,/\,N_\mathrm{K} = 7.5\%$ when $a_{12}=-56.7\,a_0$ [panel\,(e)] and $\Delta N_\mathrm{K}\,/\,N_\mathrm{K} = 13.1\%$ when $a_{12}=+56.7\,a_0$ [panel\,(f)].

It is interesting to note that this error is smaller when the inter-species interaction is attractive compared to repulsive. This situation of an attractive inter-species interaction favors the cohesion of the two-species in domain A. When the inter-species interaction is repulsive, Rb and K have a higher tendency to separate from each other, leading to a higher number of losses of K atoms from domain A to domain B, and we see that the error $\Delta N_\mathrm{K}\,/\,N_\mathrm{K}$ increases. Additional numerical simulations (not shown) indicate that the error $\Delta N_\mathrm{K}\,/\,N_\mathrm{K}$ is mainly accumulated in the first milliseconds of expansion. This is consistent with the fact that when the clouds are very dilute, the effective strength of the inter-species interaction becomes negligible, and the model becomes close to exact. This fact is obviously important for simulating accurately long expansion times.

\subsection{Single species collimation}

\begin{figure}[t!]
\includegraphics[width=\textwidth]{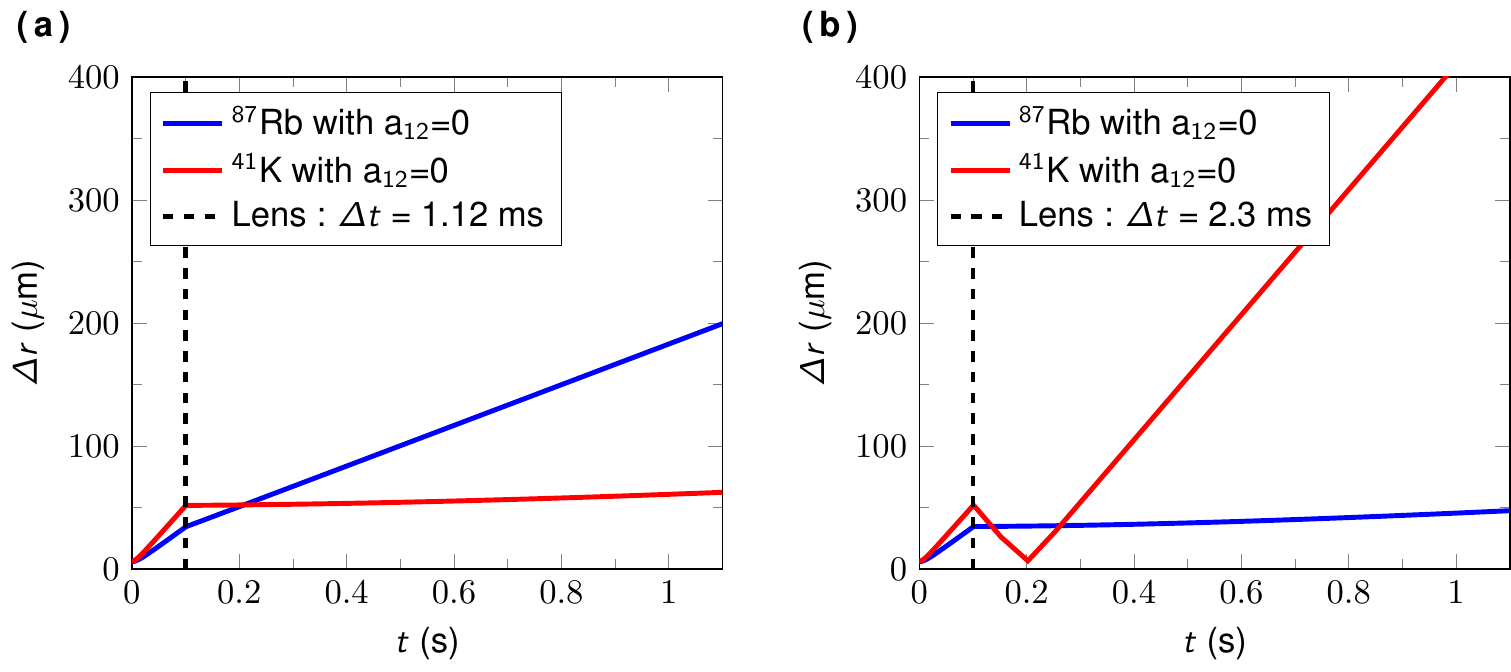}
\caption{Evolution of the characteristic sizes (standard deviation $\Delta r$) of a Rb-K mixture when $a_{12}=0$. Panel (a): the DKC pulse is optimized to collimate the K cloud. Panel (b): the DKC pulse is optimized to collimate the Rb cloud. The blue and red lines are for Rb and K, respectively. The initial trap frequencies are $\omega_{\mathrm{Rb}} = 2\pi \times 10$\,Hz and $\omega_{\mathrm{K}} = 2\pi \times 15$\,Hz.}
\label{fig_single_lens}
\end{figure}

We now consider the case where the Feshbach field is tuned such that the two BECs do not interact with each other at any time. We are interested in the kinetic expansion of the two clouds with a DKC pulse optimized to collimate one of the two species after a first free-expansion step of 100\,ms from the initial trap. Fig.\,\ref{fig_single_lens} shows the characteristic size evolution of the BECs, i.e. the standard deviations $\Delta r$ of the BECs calculated with the TD-CGPE (\ref{eq_TDCGPE}), when the lens is optimized either to collimate the K-BEC, panel (a) with a lens duration of 1.12\,ms, or to collimate the Rb-BEC, panel (b) with a lens duration of 2.3\,ms. In both cases the blue and red lines show the characteristic size evolution of the Rb and K BECs, respectively. In the first configuration [panel (a)], the lens is too short to collimate the Rb BEC. After the lens, the expansion speed of the Rb cloud is equal to 165\,$\mu$m/s, corresponding to an expansion energy of 287\,pK. In this configuration, the K cloud is well collimated and its expansion speed is only 23\,$\mu$m/s, equivalent to an energy 5.6\,pK. In the second configuration [panel (b)], the lens collimates the Rb cloud, leading to a slow expansion speed of 20\,$\mu$m/s (4.2\,pK) but it focuses the K cloud. At later times ($t \geqslant 200$\,ms) the K cloud expands at a speed of 509\,$\mu$m/s (2.7\,nK). 

In addition to the fact that the presence of a Feshbach field is not suitable for an interferometry sequence, the configuration depicted in the Fig.\,\ref{fig_single_lens} leads to a fast expansion of one of the two BECs, an effect which limits drastically the sensitivity of a dual species interferometer~\cite{Louchet_Chauvet_2011, Schubert2013, tino_precision_2013, CQGAguilera2014, ASPALTSCHUL2015}. These two configurations highlight the particular importance of the timing of the DKC pulse and the difficulty to limit the expansion speeds of the two ensembles below 100\,$\mu$m/s, as required in~\cite{ASPALTSCHUL2015} to operate at the same level than state-of-the-art classical implementations developed for testing the UFF~\cite{PRLTouboul2017}.

\subsection{Dual-species collimation with a multi-pulse atomic lenses}

\begin{figure}[t!]
\includegraphics[width=0.9\textwidth]{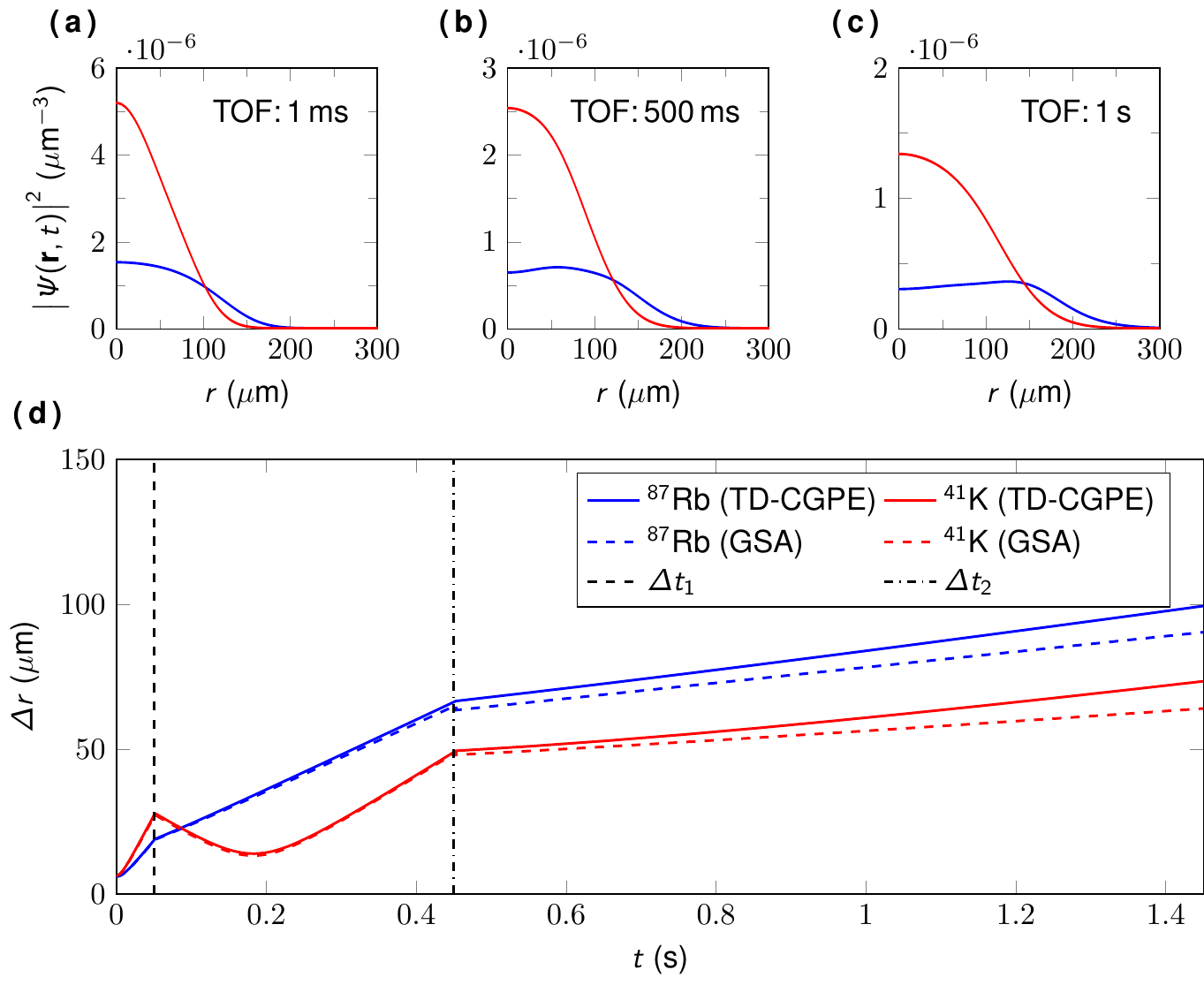}
\caption{Evolution of the densities [panels (a), (b) and (c)] and of the characteristic sizes or standard deviations $\Delta r$ [panel (d)] of a dual species BEC with a sequence of two DKC pulses. The state preparation is made of a 5-step sequence: First a free expansion during 50\,ms, followed by a DKC pulse of duration $\Delta t_1 =2.75$\,ms and by a second free expansion during 400\,ms, followed by a second DKC pulse of duration $\Delta t_2 =0.35$\,ms. This sequence takes place in presence of a Feshbach magnetic field which suppresses Rb-K interactions. The last step is a TOF where the Feshbach field is switched off 1\,ms after the second lens. The blue and red colors are for Rb and K, respectively. Panels (a-c) show representations of the TD-CGPE density distributions after the second lens for different TOFs: (a) 1\,ms, (b) 500\,ms and (c) 1\,s. In panel (d), the straight and dashed lines are  the results obtained with the TD-CGPE and with the generalized scaling approach (GSA), respectively. The two vertical dashed and dash-dotted lines mark the times at which the two DKC pulses are operating.}
\label{fig_Size_Lens_GPE_vs_SA}
\end{figure} 

To further control the dynamics of the two coupled atomic ensembles, the use of a sequence of two DKC pulses is advantageous. The strategy proposed is to prepare the two species in a trap in presence of a Feshbach resonance such that $a_{12}=0$ and to keep the Feshbach field on during all the preparation sequence. After a first release, a first DKC pulse is switched on during $\Delta t_1$ . This duration is tuned in order to slow down the expansion of the Rb cloud and to focus the K BEC. This step is followed by a second release whose duration is long enough to pass the focus point of the K cloud. At this stage, the two clouds expand in size, and a second DKC pulse of duration $\Delta t_2$ is applied to collimate both species simultaneously. The Feshbach field is then turned off 1\,ms after the last pulse, to be able to perform the interferometry sequence.

In Fig.\,\ref{fig_Size_Lens_GPE_vs_SA}(d) we show the evolution of the characteristic sizes (standard deviation) of the two BECs calculated with the TD-CGPE (straight lines) and with the generalized scaling approach (dashed lines) in the case where, after a first free expansion during 50\,ms, a first lens is applied for a duration $\Delta t_1\,=2.75$\,ms followed by a second free expansion during 400\,ms and by a second DKC pulse of duration $\Delta t_2\,=0.35$\,ms. The starting times of the two lenses are marked by vertical dashed and dash-dotted lines. The density distributions of the two species, calculated with the TD-CGPE, are shown for different TOFs after the second lens in panels (a), (b) and (c). They highlight the influence of the remaining mean-field inter-species interaction which deform progressively the density profile of the Rb cloud. After the final release, the 3D kinetic expansion speed of the Rb and K BECs, calculated with the TD-CGPE, are respectively 37.6\,$\mu$m/s and 34.6\,$\mu$m/s, corresponding to 14.9\,pK and 12.6\,pK in units of expansion energy. These expansion rates are appropriate for the most demanding high-precision dual species AI. This optimized configuration was found by scanning the large parameter space offered by the proposed strategy of using a sequences of two DKC pulses. The characteristic sizes predicted by the scaling approach during the expansion of both clouds are in good agreement with the exact calculation. This is one of the clear interest of this approach which is numerically much less demanding than solving the coupled time-dependent Gross-Pitaevskii equations and serves as a guide to effortlessly optimise the dual-lens sequence.

\begin{figure}[t!]
 \includegraphics[width=\textwidth]{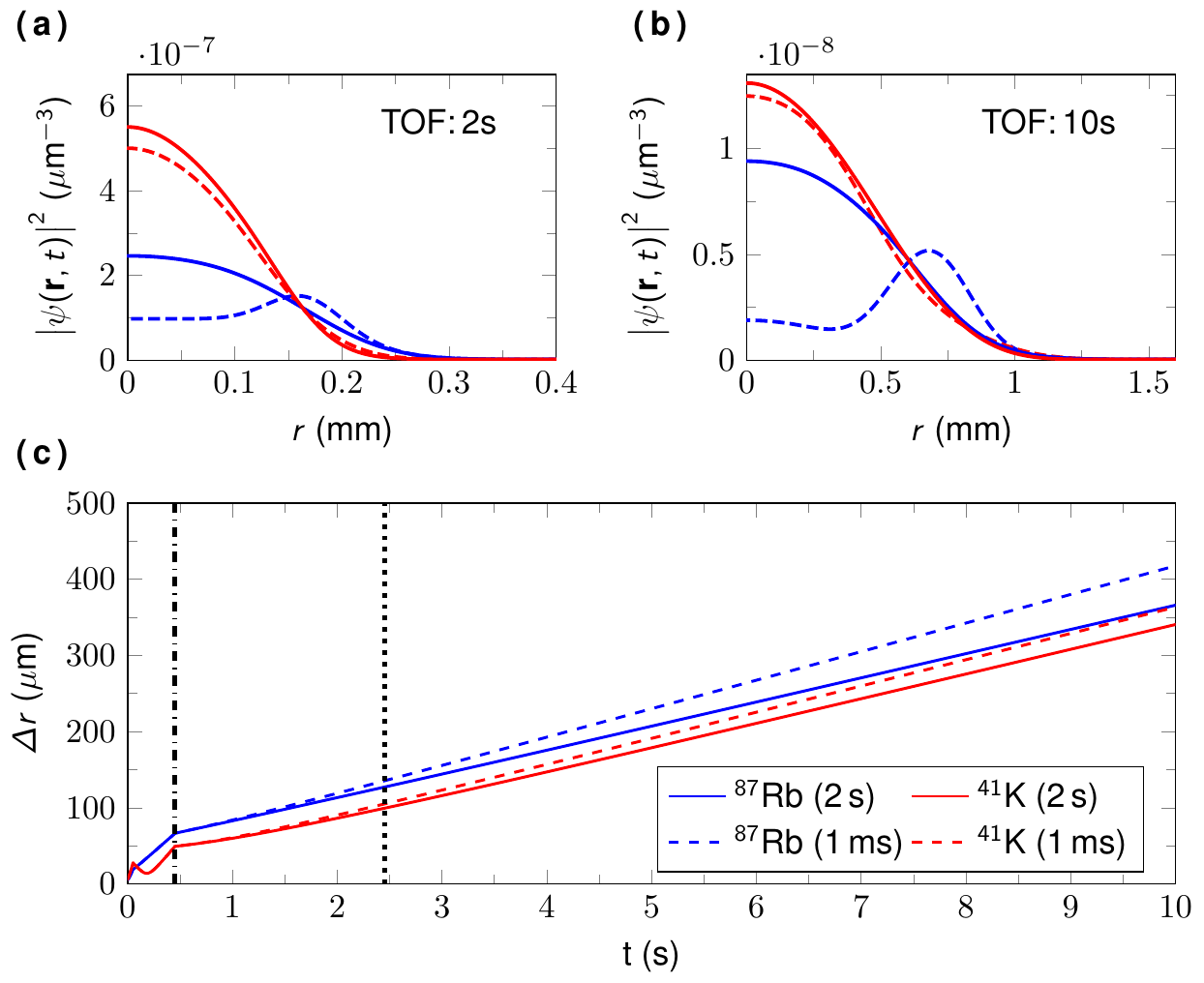}
\caption{Evolution of the densities [panels (a) and (b)] and of the characteristic sizes or standard deviations $\Delta r$ [panel (c)] of a dual species BEC with a sequence of two DKC pulses. The state preparation sequence is the same as in Fig.\,\ref{fig_Size_Lens_GPE_vs_SA}. The dashed and straight lines in the upper panels (a) and (b) denote the case where the Feshbach magnetic field is turned off respectively 1\,ms or 2\,s after the second DKC pulse. The blue and red colors denote the Rb and K species. Panel (a): Representation of the density distributions after 2\,s of TOF. Panel (b): Representation of the density distributions after 10\,s of TOF. Panel (c): Characteristic size evolution of the two BECs when the Feshbach magnetic field is turned off 2\,s after the second DKC pulse. The timings of the second DKC pulse and of the time at which the Feshbach field is switched off are marked by vertical dash-dotted and dotted lines in this panel.}
\label{fig_Size_Lens}
\end{figure}

\subsection{Impact of the inter-species mean-field interactions}\label{ssec:mfDKC}

The sequence proposed in Fig.\,\ref{fig_Size_Lens_GPE_vs_SA} has been optimized in the case where the inter-species interaction is suppressed during the preparation stage and then switched back 1\,ms after the second lens. As already mentioned, this first proposal suffers from a progressive distortion of the Rb cloud. This distortion is even more pronounced for longer TOFs, as shown by the dashed lines of panels (a) and (b) of Fig.\,\ref{fig_Size_Lens}. These two panels present in red and blue dashed lines the K and Rb density profiles after a TOF of 2 and 10\,s. This distortion arises from the fact that the inter-species mean-field interaction is not yet negligible when the Feshbach field is switched off: The residual inter-species repulsion leads to a deformation of the density distribution of the Rb cloud, which is pushed away from the central region occupied by K. Minimizing the detrimental impact of the residual inter-species mean-field repulsion on the Rb density profile requires to keep the Feshbach magnetic field for longer. Fig.\,\ref{fig_Size_Lens} depicts the case where the Feshbach magnetic field is kept for an extra 2\,s after the second DKC pulse. In panels (a) and (b) we show in solid lines the Rb and K density profiles calculated with the TD-CGPE after a TOF of 2 and 10\,s. The timings of the second DKC pulse and of the time at which the Feshbach field is switched off are marked by vertical dash-dotted and dotted lines in panel (c). In this optimized situation the shapes of the density distributions do not change and the two BECs just experience a simple size expansion. Both clouds are undistorted after 10\,s of TOF because the clouds are already so dilute when the Feshbach magnetic field is switched off that the inter-species repulsion is negligible. The kinetic expansion speeds of the Rb and K clouds are then 31.8\,$\mu$m/s and 32.6\,$\mu$m/s, corresponding to energies of 10.7\,pK and 11.2\,pK. These expansion velocities, smaller than the ones of the previous section, are even more suitable for a high precision dual species AI.

\section{Developed source concept and requirements of the UFF test}
\label{sec:high-precision}

The results of the previous section suggest the possibility of a high degree of control of the expansion rates of the two gases by exploiting the non-linear interactions and by using DKC techniques. In this section, we review systematic and statistical error sources in a test of the UFF, linked to the phase-space properties of the proposed binary source, such as wave front aberrations, mean-field fluctuations and couplings to gravity gradients and rotations. We discuss the main scaling properties and orders of magnitude involved to keep these effects below a target performance of $\delta \eta \leq 10^{-15}$ in the so-called E\"otv\"os ratio
~\cite{will_confrontation_2010} through careful interferometer input state engineering. For this assessment, we suppose a pulse separation time $T=5$\,s, an effective momentum transfer $k_i=4\times 2\pi/\lambda_i$, with $\lambda_\text{Rb}=780$\,nm and $\lambda_\text{K}=767$\,nm, and a number of atoms $N=10^6$ per shot to reach the performance goal in shot-noise-limited operations, which are typical parameters for a space-borne quantum test of the UFF along the lines of~\cite{CQGAguilera2014, NJPWilliams2016, ArxivBerge2019}.

\subsection{Excitation rates}

The efficient transfer of atoms between desired momentum states through coherent manipulation with light is essential for high-contrast interferometry. However, two-photon beam splitting mechanisms based on counter-propagating beams are Doppler-sensitive, such that velocity selection leads to spurious atoms in unwanted states affecting the signal-to-noise ratio of the interferometer. Moreover, efficient excitation requires a homogeneous beam profile over the spatial extent of the atoms. Both aspects constrain the sizes and expansion rates of the atomic ensembles, in particular in scenarios involving long drift times in the order of seconds. As an example, starting from a mm size, a Rb cloud with an effective $\mu$K expansion temperature expands up to several tens of centimeters in a few seconds, whereas for an expansion in the nK regime, the ensemble size is barely changing. Especially in space missions with limited optical power, the beam waist, and consequently the ensemble size, needs to be kept small in order to reach sufficiently high Rabi frequencies. Moreover, beam splitters based on Bragg diffraction~\cite{berman_atom_1997, meystre2001atom, mueller2008atom, Siemss2020} and Bloch oscillations
~\cite{PRLDahan1996, PRLWilkinson1996} feature relatively long interrogation times, resulting in a sharp velocity acceptance such that the velocity width of the atomic distribution typically needs to be much smaller than the recoil velocity~\cite{szigeti2012momentum}, equivalent to a few tens of nK. With DKC, these requirements are readily met as described in the previous sections of this paper as well as implemented in various experiments~\cite{PRLMuntinga2013,Mcdonald2013, ahlers2016double, Abend2016, gebbe2019twin}.

\begin{figure}[t!]
\includegraphics[width=0.9\textwidth]{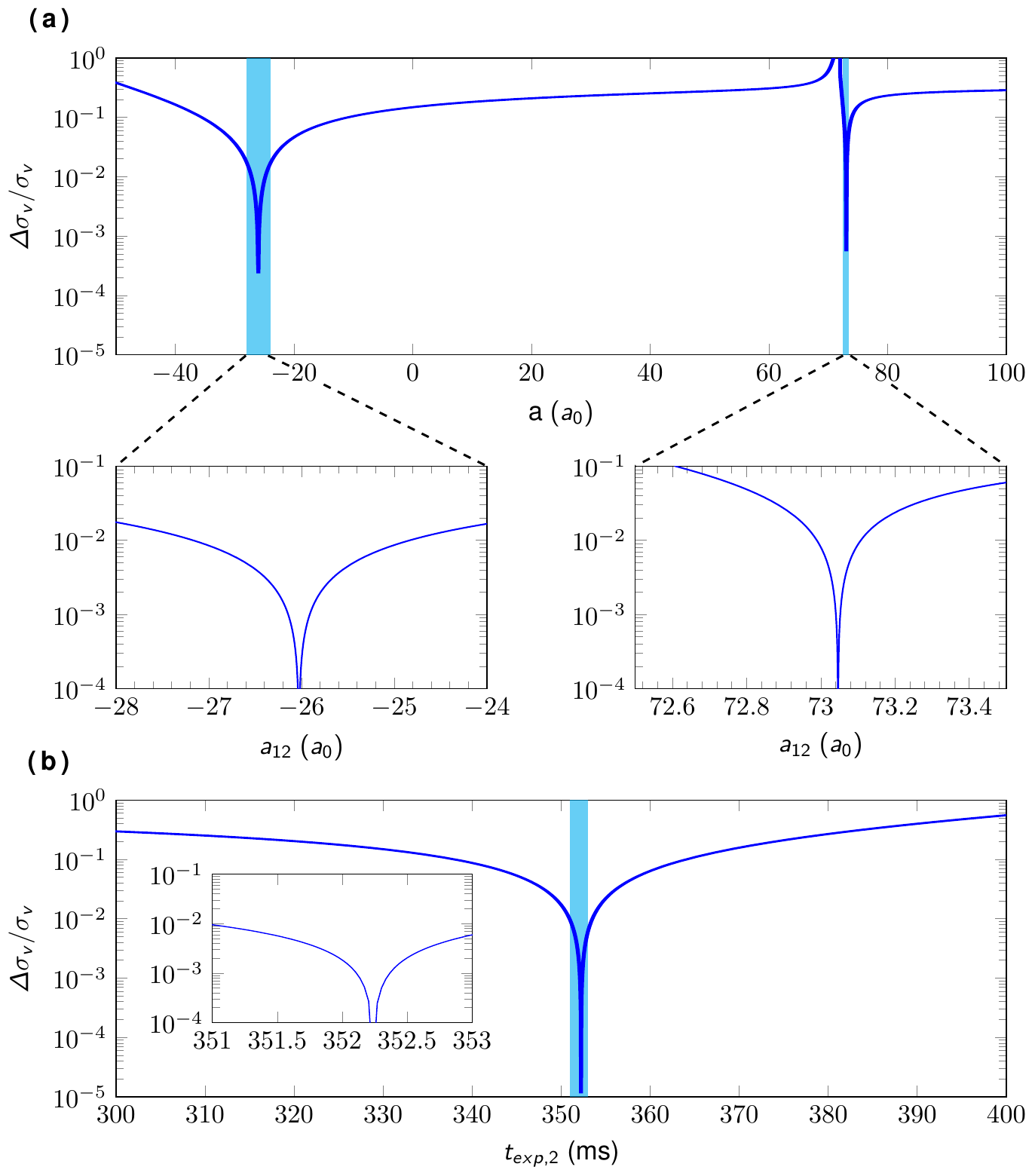}
\caption{Expansion rate matching with two different dual-pulse lens sequences. Panel\,(a): $\Delta\sigma_v/\sigma_v$ as a function of the inter-species scattering length $a_{12}$ after the second DKC pulse and during 1\,s, time at which the Feshbach field is switched off. Panel\,(b): $\Delta\sigma_v/\sigma_v$ as a function of the duration $t_{exp,2}$ of the free expansion between the two DKC pulses. The Feshbach field is turned off 1\,ms after the second DKC pulse.}
\label{fig_Delta_V_tuning}
\end{figure}

\subsection{Wave front aberrations}
Following the discussions of references,\cite{APBSchkolnik2015, Louchet_Chauvet_2011}, the wave front distortion associated with a curvature radius $R$ leads to a bias acceleration
\begin{equation}
a_{\mathrm{WF}} = \dfrac{\sigma_{v}^{2}}{R}\,,
\end{equation}
which scales with the square of the expansion rate $\sigma_v$ if a Gaussian atomic density profile is assumed. Consequently, the resulting differential acceleration of the two ensembles is determined by their relative differential expansion rate
\begin{equation}
\dfrac{\Delta\sigma_v}{\sigma_v} \equiv
\dfrac{2\left|\sigma_v^{\mathrm{Rb}}-\sigma_v^{\mathrm{K}}\right|}{\sigma_v^{\mathrm{Rb}}+\sigma_v^{\mathrm{K}}}\,.
\end{equation}
The expansion rate matching $\Delta\sigma_v/\sigma_v$ is hence the figure of merit for the mitigation of this effect and is traded-off against the curvature of the beam. For example, a $10^{-15}$ UFF test assuming a joint low expansion rate in the order of $10$\,pK requires $(\Delta\sigma_v/\sigma_v)/R$ of the order of $10^{-6}$/m.

It was shown here that scattering-free dual-pulse DKC can lead to a simultaneous reduction in the expansion rates of both species. With this technique, the inter-species scattering length $a_{12}$ is tuned to zero after the free expansion following the second lens, when the clouds are sufficiently dilute. Alternatively, $a_{12}$ can be tuned to an arbitrary non-vanishing value for a certain duration after the second lens, which can be used to manipulate the resulting differential expansion rate. The results of this mean-field assisted dual-DKC are shown in Fig.\,\ref{fig_Delta_V_tuning}(a), supposing a lens sequence as described in Sec.\,\ref{ssec:mfDKC} and switching off the Feshbach-field 1\,s after the second lens. Obviously, there are optimal values for $a_{12}$ which lead to $\Delta \sigma_v/\sigma_v \sim 10^{-3}$. However, this requires a control of the inter-species scattering length to a level better than $0.1\,a_0$, which is a challenging stability control of the Feshbach field.

A more realistic alternative illustrated in Fig.\ref{fig_Delta_V_tuning}(b) consists in controlling the duration between the two lenses followed by an immediate switch off of the Feshbach field after the second lens. Again, an optimum can be found, such that $\Delta \sigma_v/\sigma_v < 10^{-4}$ can be achieved given a control of the timing between the two lenses to a level better than 100\,$\mu$s, which is experimentally easily accessible and relaxes the requirements on the curvature to $R<100$\,m.

The promising proposed mitigation of the wave front systematic effects supposes Gaussian atomic density profiles. Deviations from that shape, as will appear for certain configurations (c.f. the dashed blue line in Fig.\,\ref{fig_Size_Lens}(b) for instance), would require a modified treatment. However, due to the large parameter space of the applied techniques (including the timings of the lenses, the durations and magnitudes of the Feshbach-fields and the possibility to include more lensing steps in the sequence), the final trade-off between overall and differential expansion rate, shape of the atomic distributions and available preparation time in an experiment should be possible in every specific case.

\subsection{Gravity gradients and rotations}

Gravity gradients $\gamma$ and rotations $\Omega$ couple to the initial position and velocity of the atoms and translates any uncertainty in their determination into an acceleration uncertainty in the interferometry measurement. Consequently, the initial center-of-mass position $r_0$ and velocity $v_0$ need to be well characterized and, in order to mitigate these systematic effects in a differential measurement, the center-of-mass overlap of the two species has to be realized to a high degree of accuracy. Thanks to recent gravity gradient compensation proposals~\cite{PRLRoura2017}, also implemented in~\cite{PRLOverstreet2017,PRLDamico2017}, the requirement on the mean position and velocity uncertainties is on the order of $\mu$m and $\mu$m/s, respectively for a UFF test at the $10
^{-15}$ level and below~\cite{Loriani2020}. For single species, this is within reach as confirmed by recent theoretical studies~\cite{NJPCorgier2018,SRAmri2019} and in line with state-of-the-art experimental realisations~\cite{Rudolph-thesis} such that the extension to binary mixtures is straightforward with the tools presented in this paper.

In a similar way, constant rotation rates, for example due to Earth's rotation, may be accounted for by counter-rotating the light-field between subsequent interferometry pulses~\cite{PRLLan2008}. However, spurious rotations couple to the center of mass velocity jitter of the atomic clouds and constrain the initial velocity mismatch to 0.3\,nm/s for typically assumed residual rotation rates in the order of $10^{-6}$\,rad/s. Verification of this control over the center-of-mass velocity, several realizations of the source preparation process are required~\cite{Schubert2013,CQGAguilera2014}. As the mean velocity uncertainty scales as $\delta v_0 = \sigma_v/\sqrt{\nu N}$ for a given expansion rate $\sigma_v$ and number of atoms $N$ per shot, a joint, low effective expansion rate of both species of 10\,pK reduces the required number of cycles to a reasonable $\nu\sim 10^4$ shots.

\subsection{Mean-field}

Variations in the mean-field energy due to atomic density fluctuations give rise to  phase noise (and hence to an acceleration error) which can be calculated by averaging over the spatial distribution and integrating over the duration $2T$ of the interferometer. In a simplified model assuming that the clouds are overlapping and not separating during the interferometry sequence, the resulting acceleration uncertainty
\begin{equation}
\delta a_i = \dfrac{3\hbar\sqrt{N_i(a_{ii}^2+a_{12}^2)}}{m_i k_i T^2}
\int_0^{2T} \dfrac{dt}{(R_{i,0}^2+\sigma_v^2 t^2)^3 }
\end{equation}
of species $i=$ K, Rb defines a minimum cloud size $R_{i,0}$ at the application of the first beam splitter for a given atom number fluctuation $\sqrt{N_i}$, intra-species (inter-species) s-wave scattering lengths $a_{ii}$ ($a_{12}$), effective expansion rate $\sigma_v$ and atomic mass $m_i$. For both species, the required cloud size is in the order of a few mm at the application of the first beam splitter, which can easily be realized by letting the ensembles expand to a sufficiently large size before lensing. In fact, this increased ratio of size-at-lens and cloud size upon release from the trap generally leads to an improved DKC performance as shown in this paper.

\section{Discussion and Perspectives}
\label{sec:conclusion}
In this paper, we presented a source concept for a dual-species interacting mixture of two quantum gases suitable to input an atom interferometer testing the universality of free fall at levels better than $10^{-15}$. The main limitation to such a test consists in the stringent requirement of observing the two gases at drift times of several seconds (about 10~s), in principle accessible to condensed gases only. We satisfy this requirement by devising a dual-delta-kick collimation stage acting as a telescope for each one of the matter waves. The engineering of such an atom optical scheme is complicated by the inter- and intra-species atomic interactions that need to be accounted for and prevent a geometric-optics-like solution. The control of these interactions is considered here by operating the atomic source close to already reported Feshbach transitions at low magnetic fields. A complete preparation sequence, alternating free expansion and DKC pulses of different durations, is found leading to an impressive compactness of the source with expansion energies of the two species in the $10$~pK regime. Optimising this sequence relies on a developed 2-species scaling approach, which validity in the relevant miscible regime is confirmed by contrasting it to the dynamics found by solving coupled-Gross-Pitaevskii equations. The compatibility of the result of our source optimisation is assessed with respect to the requirements of a beyond-state-of-the-art UFF test. Main known systematics as the wave front aberrations are mitigated taking advantage of the control over the non-linear dynamics of the degenerate clouds. Their expansion rates could, for example, be matched to the $10^{-4}$ level greatly relaxing the demand on the effective wave front curvature. Other requirements as a minimal coupling to gravity gradients or rotations, mean-field effects or the excitation rates by the interferometry pulses are checked to be fulfilling the UFF test requirement. We conclude that this source concept would be suitable for space mission proposals as STE-QUEST~\cite{CQGAguilera2014}. The same approach that we developed here and illustrated with the example of Rb-K could be generalised to any interacting quantum mixtures in a stable miscible regime. The use of anisotropic DKC traps slightly complicates the proposed scheme since more than two DKC pulses would be required. This can, however, be experimentally taken care of by a proper gauging of the external potentials forming the atomic lens.

\begin{acknowledgements}
We thank Jan-Niclas Siemß for valuable
discussions. This work is supported by the German Space Agency (DLR) with funds provided by the Federal Ministry for Economic Affairs and Energy (BMWi) due to an enactment of the German Bundestag under Grant Nos. 50WM1861 and 50WM2060, by ``Nieders\"achsisches Vorab" through the ``Quantum- and Nano-Metrology (QUANOMET)" initiative within the project QT3, through the Deutsche Forschungsgemeinschaft (DFG, German Research Foundation) under Germany's Excellence Strategy – EXC 2123 QuantumFrontiers, Project-ID 390837967, and through ”Förderung von Wissenschaft und Technik in Forschung und Lehre” for the initial funding of research in the new DLR Institutes (DLR-SI and DLR-QT). We also acknowledge support by the CRC 1227 DQmat within the projects A05 and B07, the QUEST-LFS, the Verein Deutscher Ingenieure (VDI) with funds provided by the Federal Ministry of Education and Research (BMBF) under Grant No. VDI 13N14838 (TAIOL).
RC and K. P.-T. are grateful to the German Foreign Academic Exchange (DAAD) for supporting their research activities in Germany. RC and SL acknowledge the support of the IP@Leibniz program of the Leibniz University of Hanover for travel grants supporting their stays in France. RC and NG acknowledge mobility support from the Q-SENSE project, which has received funding from the European Union's Horizon 2020 Research and Innovation Staff Exchange (RISE) Horizon 2020 program under Grant Agreement Number 691156. Additional mobility funds were thankfully made available through the bilateral exchange project PHC-Procope.
\end{acknowledgements}

\section*{References}

%

\end{document}